\documentclass[%
 reprint,
 amsmath,amssymb,
 aps,
]{revtex4-2}

\usepackage{amsfonts}
\usepackage{multirow}
\usepackage{mathrsfs}
\usepackage{graphicx}
\usepackage{amsmath}
\usepackage{amssymb}
\usepackage{bm}
\usepackage{bbm}
\usepackage{color}
\usepackage{slashed}
\usepackage{diagbox}
\usepackage{slashbox}
\usepackage{braket}
\usepackage{wrapfig}
\usepackage[hidelinks]{hyperref}
\usepackage{cleveref}
\usepackage{fancyhdr}
\usepackage{subcaption}
\numberwithin{equation}{section}
\newcommand{\calF}{\mathcal{F}}
\newcommand{\calG}{\mathcal{G}}

\begin{document}

\title[]{A Representation Transformation of Parametric Feynman Integrals}

\author{Wen Chen}
 \email{chenwenphy@gmail.com}
\affiliation{ Key Laboratory of Atomic and Subatomic Structure and Quantum Control (MOE), Guangdong Basic Research Center of Excellence for Structure and Fundamental Interactions of Matter, Institute of Quantum Matter, South China Normal University, Guangzhou 510006, China
}
\affiliation{Guangdong-Hong Kong Joint Laboratory of Quantum Matter, Guangdong Provincial Key Laboratory of Nuclear Science, Southern Nuclear Science Computing Center, South China Normal University, Guangzhou 510006, China}
\affiliation{School of Physics, Zhejiang University, Hangzhou, Zhejiang 310027, China}
\date{\today}

\begin{abstract}

A transformation on homogeneous polynomials is proposed, which is further applied to parametric Feynman integrals. The two representations related through this transformation are dual to each other. It naturally leads to dualities of Landau equations and linear integral relations between the two representations. For integrals with momentum-space correspondences, the dual representation is equivalent to the Baikov representation.

\end{abstract}

\maketitle


\section{Introduction}

Feynman integrals are a crucial element of perturbative quantum field theory.  There are several representations of Feynman integrals, with the most well-known being the Feynman-parameter representation~\cite{Feynman:1949zx,Nambu:1957shl,Nakanishi:1957,Symanzik:1958}. Variants of the Feynman-parameter representation include the Lee-Pomeransky representation~\cite{Lee2013,Lee:2014tja} and the version used by the author in refs.~\cite{Chen:2019mqc,Chen:2019fzm,Chen:2020wsh,Chen:2024xwt}. The parametric representation is powerful in the sense that it allows for both direct integration~\cite{Hepp:1966eg,Binoth:2000ps,Bogner:2007cr,Brown:2009ta,Panzer:2014caa} and integral reduction~\cite{Tarasov:1996br,Lee:2014tja,Bitoun:2017nre,Chen:2019mqc,Chen:2019fzm,Chen:2020wsh,Artico:2023jrc,Lu:2024dsb,Wang:2024hsm}, and Landau equations~\cite{Landau:1959fi,Nakanishi:1959aa,Coleman:1965xm} can easily be formulated within this framework~(see e.g. ref.~\cite{Weinzierl:2022eaz}). Furthermore, the parametric representation can reveal some hidden mathematical structures of Feynman integrals due to its connections with hypergeometric functions~\cite{Regge:1968rhi,Ponzano:1969tk,Ponzano:1970ch,Regge:1972ns,Barucchi:1973zm,Barucchi:1974bf,Gelfand:1989a,Nasrollahpoursamami:2016lhd,Vanhove:2018mto,delaCruz:2019skx,Klausen:2019hrg}, periods~\cite{Bloch:2005bh,Bogner:2007mn,Brown:2009ta}, the coaction~\cite{Abreu:2014cla,Brown:2015fyf,Abreu:2017ptx,Abreu:2021vhb,Britto:2023rig}, etc.

Another widely used representation is the Baikov representation~\cite{Cutkosky:1960sp,Baikov:1996rk,Baikov:1996iu,Baikov:2005nv,Lee:2009dh} and its variants~\cite{Frellesvig:2017aai,Chen:2022lzr,Frellesvig:2024ymq}. Although it is less suited for direct integration, the Baikov representation has the advantage that implementing cuts is straightforward~\cite{Frellesvig:2017aai,Bosma:2017ens,Harley:2017qut}, and it is more compatible with twisted cohomology theory~\cite{Mastrolia:2018uzb,Frellesvig:2019kgj,Frellesvig:2019uqt,Frellesvig:2020qot}. The Baikov representation has been extensively applied to integral reduction~\cite{Larsen:2015ped,Bohm:2017qme,Mastrolia:2018uzb,Frellesvig:2019kgj,Frellesvig:2019uqt,Frellesvig:2020qot,Boehm:2020zig,Jiang:2023oyq}, constructing differential equation basis~\cite{Bosma:2017hrk,Chen:2020uyk,Dlapa:2021qsl}, Landau analysis~\cite{Jiang:2023qnl,Chen:2023kgw,Jiang:2024eaj,Caron-Huot:2024brh}, etc.

While it is evident that these two representations should be equivalent, since both are derived from the momentum representation, it was not clear how to directly derive one representation from the other. This problem is addressed in this letter. It is demonstrated that these two representations are related by a simple transformation.

This letter is organized as follows. A transformation on homogeneous polynomials is introduced in section \ref{sec:PolTransf}. The transformation is then applied to parametric Feynman integrals in section \ref{sec:ReprTransf}. It is proven that the two representations related through this transformation are dual to each other. The equivalence between the dual representation and the Baikov representation is proven in section \ref{sec:Baik}.

\section{Polynomial transformation}\label{sec:PolTransf}
We consider the following transformation on a homogeneous polynomial $\mathcal{F}(x_1,x_2,\dots,x_{n+1})$.
\begin{subequations}\label{eq:Transf}
\begin{align}
&u_i\equiv\frac{1}{\calF}\frac{\partial\calF}{\partial x_i},\label{eq:transf1}\\
&\calG(u)\equiv\left.\frac{1}{\calF}\right|_{x=x(u)}.\label{eq:transf2}
\end{align}
\end{subequations}
We assume that the degree of $\calF$ is $L+1$. Then by virtue of the homogeneity of $\mathcal{F}$, we have
\begin{equation}\label{eq:HomCond}
\sum_{i=1}^{n+1}u_ix_i=L+1.
\end{equation}
Taking the derivatives of both sides with respect to $u_i$, we get
\begin{equation*}
x_i+\sum_{j=1}^{n+1}u_j\frac{\partial x_j}{\partial u_i}=0.
\end{equation*}
On the other hand,
\begin{equation*}
\begin{split}
\frac{1}{\calG}\frac{\partial\calG}{\partial u_i}=\calF\frac{\partial}{\partial u_i}\left(\frac{1}{\calF}\right)=-\frac{1}{\calF}\sum_{j=1}^{n+1}\frac{\partial\calF}{\partial x_j}\frac{\partial x_j}{\partial u_i}=-\sum_{j=1}^{n+1}u_j\frac{\partial x_j}{\partial u_i}.
\end{split}
\end{equation*}
A combination of the above two equations leads to
\begin{equation}
x_i=\frac{1}{\calG}\frac{\partial\calG}{\partial u_i}.
\end{equation}
Thus we see that the transformation defined by \cref{eq:Transf} is reciprocal.

Generally speaking, the transformation defined by \cref{eq:transf1} may not be reversible. Even if it is, the obtained function $\calG(u)$ may not be a rational function. However, for the $\mathcal{F}$ polynomial of a loop integral, $\calG$ is indeed a rational function, as will be clear in the subsequent sections.

\section{Representation transformation}\label{sec:ReprTransf}
As was shown in refs.~\cite{Chen:2019mqc,Chen:2019fzm,Chen:2020wsh}, a Feynman integral can be parametrized by scalar integrals of the following form.
\begin{equation}\label{eq:ParInt}
\begin{split}
&I_x(\lambda_0,\lambda_1,\ldots,\lambda_n)=\frac{\Gamma(-\lambda_0)}{\prod_{i=m+1}^{n+1}\Gamma(\lambda_i+1)}\\
&\times\int \mathrm{d}\Pi_x^{(n+1)}\calF^{\lambda_0}\prod_{i=1}^{m}x_i^{-\lambda_i-1}\prod_{i=m+1}^{n+1}x_i^{\lambda_i}~.
\end{split}
\end{equation}
Here, the integration measure $\mathrm{d}\Pi_x^{(n+1)}$ is defined by $\mathrm{d}\Pi_x^{(n+1)}=\prod_{i=1}^{n+1}\mathrm{d}x_i~\delta(1-\mathcal{E}^{(1)}(x))$, with $\mathcal{E}^{(i)}(x)$ a positive definite homogeneous function of $x$ of degree $i$. The region of integration for $x_i$ is $[0,~\infty)$ if $i>m$ and $(-\infty,~\infty)$ if $i\leqslant m$. $\calF$ is a homogeneous polynomial of the Feynman parameters $x_i$. For a parametric integral with a momentum space correspondence, the $\mathcal{F}$ polynomial is related to the well-known Symanzik polynomials $U$ and $F$ through $\mathcal{F}=F+Ux_{n+1}$, and $x_i$ with $i\leq m$ correspond to cut propagators~\cite{Chen:2020wsh}. For normal loop integrals, $\lambda_0$ is related to the space time dimension $d$ through $\lambda_0=-\frac{d}{2}$, but here it is just a parameter.
%

We introduce the function~\cite{Chen:2020wsh}
\begin{equation}\label{eq:wDef}
w_\lambda(x)\equiv e^{\frac{\lambda+1}{2}i\pi}\int_{-\infty}^{\infty}\mathrm{d}v~\frac{1}{(v+i0^+)^{\lambda+1}}e^{-ivx}~.
\end{equation}
We have
\begin{align*}
w_{-1}(x)=&~2\pi\delta(x)~,\\
w_\lambda(x)=&~2\pi\theta(x)\frac{x^\lambda}{\Gamma(\lambda+1)},~\lambda\geq 0~.\\
\end{align*}
The Fourier transform of $w_\lambda(x)$ gives
\begin{equation}\label{eq:wFourTransf}
\frac{1}{x^{\lambda+1}}=e^{\frac{\lambda+1}{2}i\pi}\int_{-\infty}^{\infty}\frac{\mathrm{d}v}{2\pi}~w_\lambda(v)e^{-ivx},~\mathrm{Im}\{x\}<0~.
\end{equation}
Obviously we can write $\frac{1}{\Gamma(\lambda+1)}\int_0^\infty\mathrm{d}xx^\lambda$ as $\frac{1}{2\pi}\int_{-\infty}^\infty\mathrm{d}xw_\lambda(x)$. By expressing $w_{\lambda}(x)$ (or $x^{-\lambda-1}$) in terms of $v^{-\lambda-1}$ [or $w_{\lambda}(v)$] by using eqs.~(\ref{eq:wDef}~and~\ref{eq:wFourTransf}), the integral $I_x$ in eq.~(\ref{eq:ParInt}) becomes
%
\begin{equation}\label{eq:xint2uint2}
\begin{split}
&I_x(\lambda_0,\lambda_1,\dots,\lambda_n)\\
=&(-2i\pi)^{-n-1}\Gamma(-\lambda_0)\exp\left(\frac{i\pi}{2}\sum_{i=1}^{n+1}\lambda_i\right)\\
&\times\int_{-\infty}^\infty\prod_{i=1}^{n+1}\mathrm{d}v_i\prod_{i=1}^mw_{\lambda_i}(v_i)\prod_{i=m+1}^{n+1}v_i^{-\lambda_i-1}\\
&\times\int_{-\infty}^\infty\mathrm{d}\Pi_x^{(n+1)}\calF^{\lambda_0}\exp\left(-i\sum_{i=1}^{n+1}v_ix_i\right)\\
=&(-2i\pi)^{-n-1}\Gamma(-\lambda_0)\exp\left(\frac{i\pi}{2}\sum_{i=1}^{n+1}\lambda_i\right)\\
&\times\int_{-\infty}^\infty\mathrm{d}\Pi_v^{(n+1)}\prod_{i=1}^mw_{\lambda_i}(v_i)\prod_{i=m+1}^{n+1}v_i^{-\lambda_i-1}\\
&\times\int_{-\infty}^\infty\prod_{i=1}^{n+1}\mathrm{d}x_i~\calF^{\lambda_0}\exp\left(-i\sum_{i=1}^{n+1}v_ix_i\right)~,
\end{split}
\end{equation}
%
where $\mathrm{d}\Pi_v^{(n+1)}\equiv\prod_{i=1}^{n+1}\mathrm{d}v_i~\delta(1-\mathcal{E}^{(1)}(v))$. The last step of this equation can be proven by inserting an integral $\int\mathrm{d}s\delta(s-\mathcal{E}^{(1)}(v))$, rescaling $x$ and $v$ by $x_i\to\frac{x_i}{s}$ and $v_i\to sv_i$, and finally integrating out $s$.

We consider a special class of integrals of which the $\calF$ polynomials are complete. Here, by complete, we mean that there are $n$ linearly independent matrices $\Lambda_{i,jk}$ such that
\begin{equation}\label{eq:ComplDef}
\sum_{j,k=1}^{n+1}\Lambda_{i,jk}u_jx_k=0,~i=1,2,\dots,n~.
\end{equation}
We further assume that
\begin{subequations}\label{eq:TracCond}
\begin{align}
\Lambda_{i,j(n+1)}=&~0,~j\neq n+1~,\\
\sum_{j=1}^{n}\Lambda_{i,jj}=&~n_0\Lambda_{i,(n+1)(n+1)}~,
\end{align}
\end{subequations}
with $n_0$ a constant. As can be seen in \cref{app:proof2}, a family of loop integrals with a complete set of propagators do satisfy these conditions. That is, the completeness of propagators implies the completeness of the corresponding $\calF$ polynomial. The first condition of eqs.~(\ref{eq:TracCond}) is a consequence of the fact that $\calF$ is linear in $x_{n+1}$. For a complete homogeneous $\calF$ polynomial satisfying the conditions in eqs.~(\ref{eq:TracCond}), it can be proven that (see \cref{app:proof1})
\begin{equation}\label{eq:FInt2GInt}
\begin{split}
&\int_{-\infty}^\infty\prod_{i=1}^{n+1}\mathrm{d}x_i~\calF^{\lambda_0}\exp\left(i\sum_{i=1}^{n+1}v_ix_i\right)\\
=&~(-2i\pi)^{n+1}e^{\frac{i}{2}\pi(L+1)\lambda_0}C_\calF v_{n+1}^{-n_0-1}\calG(v)^{\lambda_0^\prime}~,
\end{split}
\end{equation}
where $\lambda_0^\prime\equiv-\lambda_0-\frac{n-n_0}{L+1}$ and $C_\calF$ is a ($v$-independent) constant. Substituting eq.~(\ref{eq:FInt2GInt}) into eq.~(\ref{eq:xint2uint2}) gives
\begin{equation}\label{eq:xint2uint}
\begin{split}
&I_x(\lambda_0,\lambda_1,\dots,\lambda_n)\\
=&\exp\left(\frac{i\pi}{2}\left[(L+1)\lambda_0+\sum_{i=1}^{n+1}\lambda_i\right]\right)\\
&\times C_\calF\frac{\Gamma(-\lambda_0)}{\Gamma(-\lambda_0^\prime)} ~I_u(\lambda_0^\prime,\lambda_1^\prime,\dots,\lambda_n^\prime)~,
\end{split}
\end{equation}
where $\lambda_{n+1}^\prime=\lambda_{n+1}+n_0+1$, and $\lambda_i^\prime=\lambda_i,~i\neq 0,n+1$. The integral $I_u$ is defined by
\begin{align}\label{eq:DualIntDef}
&I_u(\lambda_0^\prime,\lambda_1^\prime,\dots,\lambda_n^\prime)\\
\equiv&~\Gamma(-\lambda_0^\prime)\int_{-\infty}^\infty\mathrm{d}\Pi_u^{(n+1)}\calG(u)^{\lambda_0^\prime}\prod_{i=1}^mw_{\lambda_i^\prime}(u_i)\prod_{i=m+1}^{n+1}u_i^{-\lambda_i^\prime-1}~.\nonumber
\end{align}

Thus, we have proven that $I_x(\lambda_0,\lambda_1,\dots,\lambda_n)$ equals $I_u(\lambda_0^\prime,\lambda_1^\prime,\dots,\lambda_n^\prime)$ up to some constants, and there is an elegant duality between these two representations, which is summarized in \cref{tab:Dual}.
\begin{widetext}
\begin{center}
\begin{table}[h]
\begin{tabular}{|c|c|}
\hline
~&~\\
$\calG=\frac{1}{\calF}$ & $\calF=\frac{1}{\calG}$\\
~&~\\
\hline
~&~\\
$u=\frac{1}{\calF}\frac{\partial\calF}{\partial x}$ & $x=\frac{1}{\calG}\frac{\partial\calG}{\partial u}$\\
~&~\\
\hline
~&~\\
$\frac{1}{u^{\lambda+1}}=e^{-\frac{\lambda+1}{2}i\pi}\int_{-\infty}^{\infty}\frac{\mathrm{d}x}{2\pi}~w_\lambda(x)e^{iux}$ ~&~$~w_\lambda(x)= e^{\frac{\lambda+1}{2}i\pi}\int_{-\infty}^{\infty}\mathrm{d}u~\frac{1}{u^{\lambda+1}}e^{-iux}$~\\
~&~\\
\hline
\end{tabular}
\caption{The duality between the two representations. Here the $\frac{1}{u^{\lambda+1}}$ should be understood as $\frac{1}{(u+i0^+)^{\lambda+1}}$.}\label{tab:Dual}
\end{table}
\end{center}
\end{widetext}

\section{The Baikov representation}\label{sec:Baik}
We consider a loop integral with propagators
\begin{equation}\label{eq:PropDef}
v_i=\sum_{j,k}A_{i,jk}l_j\cdot l_k+2\sum_{j,k} B_{i,jk}l_j\cdot p_k+C_i,\quad i=1,~2,\dots,n~,
\end{equation}
where $l$ are the loop momenta and $p$ are the external momenta. For simplicity, we assume that the Gram determinant of the external momenta $p$ does not vanish, and all the propagators are complete and linearly independent. Here, by complete, we mean that $v_i$ form a complete basis such that all the scalar products $l_i\cdot l_j$ and $l_i\cdot p_j$ can be expressed in terms of $v_i$. For future convenience, we collectively denote $l$ and $p$ by $q$. Scalar integrals can be parametrized by integrals of the form~\cite{Baikov:1996rk,Baikov:1996iu}
\begin{equation}\label{eq:Baik}
\begin{split}
&J(\lambda_0,\lambda_1,\ldots,\lambda_n)=\\
&\int\prod_{i=1}^n\mathrm{d}v_n~P^{-\lambda_0-\frac{1}{2}(L+E+1)}\prod_{i=1}^nv_i^{-\lambda_i-1}~,
\end{split}
\end{equation}
where $E$ is the number of the external momenta $p$. Here we have omitted some irrelevant prefactors for brevity. Notice that we use such a convention that the indices of the propagators are $-\lambda_i-1$ rather than the more conventional $-\lambda_i$. The polynomial $P(v)$ is proportional to the Gram determinant of the momenta $q$. That is,
\begin{equation}
P(v)\equiv\frac{\det(q_i\cdot q_j)}{\det(p_i\cdot p_j)}~.
\end{equation}
In this section, we will prove that $J(\lambda_0,\lambda_1,\ldots,\lambda_n)$ is equivalent to the integral $I_u(\lambda_0^\prime,\lambda_1^\prime,\dots,\lambda_n^\prime)$ in \cref{eq:DualIntDef} up to some constants.

Taking $\mathcal{E}^{(1)}(u)=|u_{n+1}|$ and $m=0$ in the integral $I_u$, we get
\begin{align*}
&I_u(\lambda_0^\prime,\lambda_1^\prime,\dots,\lambda_n^\prime)\\
=&\Gamma(-\lambda_0^\prime)\int_{-\infty}^\infty\prod_{i=1}^n\mathrm{d}u_i~\left[\calG_+^{\lambda_0^\prime}+\calG_-^{\lambda_0^\prime}e^{-i\pi(\lambda_{n+1}^\prime+1)}\right]\prod_{i=1}^{n}u_i^{-\lambda_i^\prime-1}\\
=&\Gamma(-\lambda_0^\prime)\int_{-\infty}^\infty\prod_{i=1}^n\mathrm{d}u_i~\prod_{i=1}^{n}u_i^{-\lambda_i^\prime-1}\\
&\times\left[\calG_-^{\lambda_0^\prime}e^{i\pi(\lambda_{n+1}^\prime+1)}+\calG_-^{\lambda_0^\prime}e^{-i\pi(\lambda_{n+1}^\prime+1)}\right]\\
=&2\Gamma(-\lambda_0^\prime)\cos\left(\pi(\lambda_{n+1}^\prime+1)\right)\int_{-\infty}^\infty\prod_{i=1}^n\mathrm{d}u_i~\calG_-^{\lambda_0^\prime}\prod_{i=1}^{n}u_i^{-\lambda_i^\prime-1}\\
=&\frac{-4i\pi}{\Gamma(1+\lambda_0^\prime)}\cos\left(\pi\lambda_{n+1}^\prime\right)\\
&\times\int_{-\infty}^\infty\prod_{i=1}^n\mathrm{d}u_i~\theta\left(-\calG_-\right)\left(-\calG_-\right)^{\lambda_0^\prime}\prod_{i=1}^{n}u_i^{-\lambda_i^\prime-1}~,
\end{align*}
where $\calG_+\equiv\left.\calG\right|_{u_{n+1}=1}$ and $\calG_-\equiv\left.\calG\right|_{u_{n+1}=-1}$. Due to the positivity of $|u_{n+1}|$, the integral receives contributions from both $u_{n+1}=1$ and $u_{n+1}=-1$. The second step of this equation is carried out by replacing the integration variables with $x_i\to -x_i$ for the first term in the square bracket. In the last step, we have closed the contour around the branch cut. As a result, only the discontinuity of $\calG_-^{\lambda_0^\prime}$ contributes to the integral. The discontinuity arises from the region where $\calG_-<0$, which gives rise to the theta function $\theta\left(-\calG_-\right)$. Notice that we use such a convention that $\mathrm{Im}\{\calF\}<0$. Thus $\mathrm{Im}\{\calG\}>0$. In the last step we have expressed $\sin(\pi\lambda_0^\prime)$ in terms of $\frac{-\pi}{\Gamma(-\lambda_0^\prime)\Gamma(1+\lambda_0^\prime)}$. Since $n_0=n-\frac{1}{2}(L+1)(L+E+1)$ (see \cref{app:proof2}), we have $\lambda_0^\prime=-\lambda_0-\frac{1}{2}(L+E+1)$. Comparing the above integral with the Baikov integral in \cref{eq:Baik}, we see that these two integrals are equivalent if
\begin{equation}\label{eq:G2P}
\begin{split}
P(v_1,v_2,\dots,v_n)=&-\calG_-\\
=&-\calG(v_1,v_2,\dots,v_n,-1)~.
\end{split}
\end{equation}
Immediately, we will see that this is indeed true.

We define
\begin{align*}
A_{ij}\equiv&\sum_{k=1}^nA_{k,ij}x_k~,\\
B_{ij}\equiv&\sum_{k=1}^nB_{k,ij}x_k~,\\
C\equiv&\sum_{i=1}^nC_ix_i~.
\end{align*}
Then we have~\cite{Chen:2019mqc}
\begin{equation}\label{eq:Fforloop}
\calF=\sum_{i,j,k,l}p_k\cdot p_l\tilde{A}_{ij}B_{ik}B_{jl}+(x_{n+1}-C)U~,
\end{equation}
where
\begin{align*}
U\equiv&\det(A_{ij})~,\\
\tilde{A}_{ij}\equiv&\left(A^{-1}\right)_{ij}U~.
\end{align*}
Let $a_{ij}$, $b_{ij}$, $\alpha_{ij,k}$, and $\beta_{ij}$ be those defined in ref.~\cite{Chen:2019fzm}. That is, they are solutions to the following equations.
\begin{subequations}\label{eq:AlphBetDef}
\begin{align}
&\sum_la_{il}B_{l,jk}=0,\\
&\sum_lb_{il}A_{l,jk}=0,\\
&\sum_{kl}\alpha_{ij,k}a_{kl}A_{l,mn}=\frac{1}{2}\left(\delta^{im}\delta^{jn}+\delta^{in}\delta^{jm}\right),\\
&\sum_{kl}\beta_{ij,k}b_{kl}B_{l,mn}=\delta^{im}\delta^{jn}~.
\end{align}
\end{subequations}
Then it can be proven that (see \cref{app:proof2})
\begin{equation}\label{eq:Finu}
\begin{split}
\calG=~&u_{n+1}\det\left[\sum_{kl}\alpha_{ij,k}a_{kl}(u_l+u_{n+1}C_l)\right.\\
&\left.+ u_{n+1}\sum_{mn}g_{mn}B^\prime_{im}B^\prime_{jn}\right]~,
\end{split}
\end{equation}
where $u_i$ are those appeared in \cref{eq:Transf}, $g_{ij}$ is the inverse of the Gram matrix $p_i\cdot p_j$, and
\begin{equation*}
B^\prime_{ij}(u)\equiv\frac{1}{2u_{n+1}}\sum_{k,l}\beta_{ij,k}b_{kl}\left(u_l+C_lu_{n+1}\right)~.
\end{equation*}

On the other hand, $q_i\cdot q_j$ can be understood as a block matrix. Then we have
\begin{align*}
P=~&\frac{1}{\det(p_i\cdot p_j)}\det\begin{pmatrix}
l_i\cdot l_j&l_i\cdot p_j\\
l_i\cdot p_j&p_i\cdot p_j
\end{pmatrix}\\
=~&\det\left(l_i\cdot l_j-\sum_{mn}g_{mn}l_i\cdot p_ml_j\cdot p_n\right)~.
\end{align*}
Solving eq.~(\ref{eq:PropDef}) for $l_i\cdot l_j$ and $l_i\cdot p_j$, we get
\begin{align*}
l_i\cdot p_j=&\frac{1}{2}\sum_{k,l}\beta_{ij,k}b_{kl}(v_l-C_l)\equiv-B^{\prime\prime}_{ij}=-\left.B^{\prime}_{ij}(v)\right|_{v_{n+1}=-1}~,\\
l_i\cdot l_j=&\sum_{k,l}\alpha_{ij,k}a_{kl}(v_l-C_l)~.
\end{align*}
Thus
\begin{equation}\label{eq:Pinu}
\begin{split}
P=&\det\left(\sum_{kl}\alpha_{ij,k}a_{kl}(v_l-C_l)\right.\\
&\left.-\sum_{mn}g_{mn}B^{\prime\prime}_{im}B^{\prime\prime}_{jn}\right)~.
\end{split}
\end{equation}
Comparing \cref{eq:Finu} with \cref{eq:Pinu}, we get \cref{eq:G2P}.

\section{Discussion}\label{sec:disc}
A reversible transformation on homogeneous polynomials is proposed. The transformation is further applied to parametric Feynman integrals, since a parametric integral is characterized by a homogeneous polynomial. An exact duality between the two representations related through this transformation is proven, as summarized in \cref{tab:Dual}. Furthermore, it is proven that the dual representation is equivalent to the Baikov representation.

The transformation and the corresponding duality can be widely applied to the studies of Feynman integrals, including integral reduction and Laundau analysis. Specifically, \cref{eq:ComplDef} establishes a correspondence between a syzygy equation in the Baikov representation~\cite{Larsen:2015ped,Bohm:2017qme} and a parametric annihilator in the parametric representation~\cite{Lee:2014tja,Bitoun:2017nre}. Additionally, the transformation in \ref{eq:Transf} establishes a duality between the first (second) Landau equation in the parametric representation and the second (first) Landau equation in the Baikov representation. It is known that the infrared singularities of Feynman integrals can be identified with solutions of Landau equations. The method of regions~\cite{Beneke:1997zp,Smirnov:1999bza,Smirnov:2002pj,Pak:2010pt,Jantzen:2011nz} provides a systematic method to isolate the infrared structures of Feynman integrals. While endpoint singularities (in the parametric space) can be identified using a geometric method~\cite{Pak:2010pt,Gardi:2022khw,Ma:2023hrt}, this method can not directly be applied to more general regions, including threshold regions~\cite{Beneke:1997zp,Jantzen:2012mw,Ananthanarayan:2018tog,Gardi:2024axt}. The latter are related to solutions of the second Landau equations which are away from the origin. By virtue of the duality found in this letter, it is expected that this kind of regions can be identified by applying the geometric method to the dual representation. This is confirmed through some one-loop integrals, including those considered in refs.~\cite{Beneke:1997zp,Jantzen:2012mw}. The application of this method to multiloop integrals is complicated by the presence of irreducible scalar products, which deserves further investigation.

\begin{acknowledgments}
This work is supported by the Natural Science Foundation of China (NSFC) under the contract No. 11975200, and by Guangdong Major Project of Basic and Applied Basic Research~(No. 2020B0301030008).
\end{acknowledgments}

\bibliographystyle{apsrev4-2} 
\bibliography{refs.bib}

\appendix

\section{Proof of eq.~(\ref{eq:FInt2GInt})}\label{app:proof1}
Combining the homogeneity condition \cref{eq:HomCond} with those conditions in \cref{eq:ComplDef,eq:TracCond}, we have
\begin{equation}\label{eq:ComplDef2}
\begin{split}
0=&\sum_{j,k=1}^{n+1}\Lambda_{i,jk}u_jx_k\\
=&\sum_{j=1}^{n+1}\sum_{k=1}^n\Lambda_{i,jk}u_jx_k+\Lambda_{i,(n+1)(n+1)}u_{n+1}x_{n+1}\\
=&(L+1)\Lambda_{i,(n+1)(n+1)}+\sum_{j=1}^{n+1}\sum_{k=1}^n\Lambda^\prime_{i,jk}u_jx_k~,
\end{split}
\end{equation}
with
\begin{equation*}
\Lambda^\prime_{i,jk}\equiv\Lambda_{i,jk}-\Lambda_{i,(n+1)(n+1)}\delta_{jk}~.
\end{equation*}
For simplicity, we denote
\begin{equation*}
I_0(f)\equiv\int_{-\infty}^\infty\prod_{i=1}^{n+1}\mathrm{d}x_i~\calF^{\lambda_0}f(x)\exp\left(-i\sum_{i=1}^{n+1}v_ix_i\right)~,
\end{equation*}
with $f(x)$ a polynomial in $x$. Then, by virtue of eq.~(\ref{eq:ComplDef2}), we have
\begin{align*}
&-(L+1)\lambda_0\Lambda_{i,(n+1)(n+1)}I_0(1)\\
=&\lambda_0\sum_{j=1}^{n+1}\sum_{k=1}^n\Lambda^\prime_{i,jk}I_0(u_jx_k)\\
=&\sum_{j=1}^{n+1}\sum_{k=1}^n\Lambda^\prime_{i,jk}\int_{-\infty}^\infty\prod_{i=1}^{n+1}\mathrm{d}x_i~\frac{\partial\calF^{\lambda_0}}{\partial x_j}x_k\exp\left(-i\sum_{i=1}^{n+1}v_ix_i\right)\\
=&-\sum_{j=1}^{n}\Lambda^\prime_{i,jj}I_0(1)+i\sum_{j=1}^{n+1}\sum_{k=1}^n\Lambda^\prime_{i,jk}v_jI_0(x_k)\\
=&(n-n_0)\Lambda_{i,(n+1)(n+1)}I_0(1)-\sum_{j=1}^{n+1}\sum_{k=1}^nv_j\frac{\partial I_0(1)}{\partial v_k}~.
\end{align*}
That is
\begin{align*}
\sum_{j=1}^{n+1}\sum_{k=1}^n\Lambda^\prime_{i,jk}v_j\frac{\partial I_0(1)}{\partial v_k}=&\left[n-n_0+(L+1)\lambda_0\right]\\
&\times\Lambda_{i,(n+1)(n+1)}I_0(1)~.
\end{align*}
On the other hand, eq.~(\ref{eq:ComplDef2}) is equivalent to
\begin{equation*}
\sum_{j=1}^{n+1}\sum_{k=1}^n\Lambda^\prime_{i,jk}v_j\frac{\partial\calG(v)}{\partial v_k}=-(L+1)\Lambda_{i,(n+1)(n+1)}\calG~.
\end{equation*}
Thus we get
\begin{equation*}
\sum_{j=1}^{n+1}\sum_{k=1}^n\Lambda^\prime_{i,jk}v_j\frac{\partial}{\partial v_k}\left[\calG(v)^{\lambda_0+\frac{n-n_0}{L+1}}I_0(1)\right]=0~.
\end{equation*}
By assumption, the matrix $\Lambda^\prime_{ij}(v)\equiv\sum_{k=1}^{n+1}\Lambda^\prime_{i,kj}v_k$ is invertible. Thus $\calG(v)^{\lambda_0+\frac{n-n_0}{L+1}}I_0(1)$ is independent of $v_i,~i\neq n+1$. That is
\begin{equation}
\begin{split}
I_0(1)=&(-2i\pi)^{n+1}e^{\frac{i}{2}\pi(L+1)\lambda_0}C_\calF\\
&\times v_{n+1}^{-n_0-1}\calG(v)^{-\lambda_0-\frac{n-n_0}{L+1}}~,
\end{split}
\end{equation}
where $C_\calF$ is a constant, and the factors in $i$ and $\pi$ are added for future convenience. The $v_{n+1}$ dependence is fixed through a simple power counting.

\section{Derivation of eq.~(\ref{eq:Finu})}\label{app:proof2}
By virtue of \cref{eq:Fforloop}, we have (in this section, summations over repeated indices are always implied.)
\begin{align*}
\beta_{ij,k}b_{kl}u_l\calF=&\beta_{ij,k}b_{kl}\frac{\partial\calF}{\partial x_l}\\
=&2\tilde{A}_{ik}B_{kl}p_l\cdot p_j-\beta_{ij,k}b_{kl}C_lU\\
=&2\left(A^{-1}\right)_{ik}B_{kl}p_l\cdot p_ju_{n+1}\calF\\
&-\beta_{ij,k}b_{kl}C_lu_{n+1}\calF~.
\end{align*}
Multiplying both sides with $\frac{1}{2u_{n+1}\calF}A_{ij}g_{kl}$, we get
\begin{equation}\label{eq:Bt2B}
B_{ij}=A_{ik}g_{jl}B^\prime_{kl}~.
\end{equation}
Similarly,
\begin{align*}
&\alpha_{ij,k}a_{kl}u_l\calF=\alpha_{ij,k}a_{kl}\frac{\partial\calF}{\partial x_l}\\
=&-U\left[p_m\cdot p_n\left(A^{-1}\right)_{ik}\left(A^{-1}\right)_{jl}B_{km}B_{ln}+\alpha_{ij,k}a_{kl}C_l\right]\\
&+\tilde{A}_{ij}\left[p_k\cdot p_l\left(A^{-1}\right)_{mn}B_{mk}B_{nl}+x_{n+1}-C\right]\\
=&-\calF u_{n+1}\left(g_{mn}B^\prime_{im}B^\prime_{jn}+\alpha_{ij,k}a_{kl}C_l\right)+\left(A^{-1}\right)_{ij}\calF~.
\end{align*}
That is
\begin{equation}\label{eq:InvAinBt}
\begin{split}
\left(A^{-1}\right)_{ij}=&\alpha_{ij,k}a_{kl}(u_l+u_{n+1}C_l)\\
&+ u_{n+1}g_{mn}B^\prime_{im}B^\prime_{jn}~.
\end{split}
\end{equation}
Taking the determinant of both sides, we get
\begin{equation}
\begin{split}
\calG=&\calF^{-1}=u_{n+1}U^{-1}=u_{n+1}\det\left(A^{-1}\right)\\
=&u_{n+1}\det\left[\alpha_{ij,k}a_{kl}(u_l+u_{n+1}C_l)\right.\\
&\left.+ u_{n+1}g_{mn}B^\prime_{im}B^\prime_{jn}\right]~.
\end{split}
\end{equation}

Following some similar algebras, we can also get
\begin{subequations}\label{eq:Syz}
\begin{align}
0=&A_{ik}\beta_{kj,l}b_{lm}(u_m+C_mu_{n+1})\nonumber\\
&-2B_{ik}p_k\cdot p_ju_{n+1}~,\label{eq:Syz1}\\
0=&(L+1)\left(A_{ik}\alpha_{kj,l}a_{lm}+\frac{1}{2}B_{ik}\beta_{jk,l}b_{lm}\right)\nonumber\\
&\times(u_m+C_mu_{n+1})-\delta_{ij}u_kx_k~.\label{eq:Syz2}
\end{align}
\end{subequations}
Here, the first equation is obtained from eq.~(\ref{eq:Bt2B}) by multiplying both sides with $u_{n+1}$, and the second equation is obtained from eq.~(\ref{eq:InvAinBt}) by multiplying both sides with the matrix $A_{ij}$ and replacing $\delta_{ij}$ by $\frac{1}{L+1}\delta_{ij}u_kx_k$. These equations are of the form of the complete condition \cref{eq:ComplDef}, since $A_{ij}$ and $B_{ij}$ are linear in $x$. It is easy to see that they contain $\frac{1}{2}L(L+1)+LE=n$ independent equations. $\sum_{j=1}^n\Lambda_{i,jj}$ can be obtained from \cref{eq:ComplDef} by replacing $u_jx_k$ by $\delta_{jk}$. Thus \cref{eq:Syz1} gives [remember the definitions in eqs.~(\ref{eq:AlphBetDef}).]
\begin{equation*}
\sum_{j=1}^n\Lambda_{i,jj}=\Lambda_{i,(n+1)(n+1)}=0~,
\end{equation*}
and \cref{eq:Syz2} gives (notice that the matrix $A_{ij}$ is of dimension $L\times L$, and the matrix $B_{ij}$ is of dimension $L\times E$.)
\begin{align*}
\Lambda_{i,(n+1)(n+1)}=&-1~,\\
\sum_{j=1}^n\Lambda_{i,jj}=&\frac{1}{2}(L+1)(L+1+E)-n~.
\end{align*}
That is,
\begin{equation}
n_0=n-\frac{1}{2}(L+1)(L+1+E)~.
\end{equation}

\end{document}